\documentclass[12pt]{article}
\begin{document}
\title{Anomalous Fermions II}
\author{B.G. Sidharth\\
International Institute for Applicable Mathematics \& Information Sciences\\
Hyderabad (India) \& Udine (Italy)\\
B.M. Birla Science Centre, Adarsh Nagar, Hyderabad - 500 063
(India)}
\date{}
\maketitle
\begin{abstract}
We briefly comment upon the parallel between graphene and high energy fermions.
\end{abstract}
\section{Introduction}
The author had pointed out starting 1995 that in two dimensions and one dimension electrons will display strange neutrino like properties \cite{r1,r2,r3}. In fact a two component equation is obeyed \cite{greiner}. This equation is
\begin{equation}
\left(\sigma^\mu \partial_\mu - \frac{mc}{\hbar}\right) \psi = 0\label{A}
\end{equation}
where $\sigma^\mu$ denote $2 \times 2$ matrices. In case the mass vanishes (\ref{A}) gives the neutrino equation.
This has relevance to graphene that was discovered nearly a decade later.
The neutrino like equation is
\begin{equation}
\nu_F \vec {\sigma} \cdot \vec{\nabla} \psi (r) = E \psi (r)\label{e1}
\end{equation}
$\nu_F \sim 10^6 m/s$ is the Fermi velocity replacing $c$, the velocity of light and $\psi (r)$  being a two component wave function  , $\vec{\sigma}$ and $E$ denoting the Pauli matrices and energy.\\
In any case Landau had shown several decases ago that such two and one dimensional structures would be unstable and as such cannot exist -- and this was proved wrong.\\
However, there is no Lorentz invariance (except in the case of a hypothetical infinite sheet) and the two component wave function $\psi (r)$ in (\ref{e1}) comes from the wave functions in two side by side
honey comb lattices. We will see this later. This is rather like spin up and spin down.
\section{The Graphene Test Bed}
To continue, we point out that graphene can be a test bed for high energy physics. Firstly (\ref{e1}) represents a neutrino like (massless) Fermion. Indeed the massless feature has been experimentally confirmed. These are quasi particles. If we consider bi-layer graphene then even the mass comes in.\\
Interestingly graphene behaves like a "chess board", that is there is a minimum "length" \cite{physrev}. So a non-commutative geometry holds.\\
In this case we have
\begin{equation}
[x_\imath , x_j ] = \Theta^{\imath j} l^2\label{B}
\end{equation}
where as can be seen the coordinates $x_\imath$ and $x_j$ do not commute. As a result of this the Maxwell equations get modified with an extra term, as shown in detail elsewhere \cite{bgshj2013,pstj}:
\begin{equation}
\partial ^{\mu } F_{\mu \nu } =\frac{4\pi }{c} j_{\nu } + A_{\lambda}\epsilon F_{\mu \nu } \label{e15a}
\end{equation}
where the symbols have their usual meaning. In the above (\ref{e15a}) $\varepsilon $ is a dimensionless number which
is equal to one for our non-commutative case namely (\ref{B}), and is zero otherwise. With $\varepsilon = 0$ we get back the usual covariant Maxwell equations. Specializing to two dimensions we get
\begin{equation}
\partial ^{1} F_{14} =\frac{4\pi }{c} j_{4} + A_{2}
\varepsilon F_{14} \label{e16}
\end{equation}
and similar equations for the $j_1$ and $j_2$. In this case, using the electromagnetic tensor we get equations like
\begin{equation}
\frac{\partial E_{x} }{\partial x} = - {4\pi } \frac{\partial
\rho }{\partial t} +\varepsilon {A_{y}}{E_{x} }\label{e23}
\end{equation}
\begin{equation}
\frac{\partial E_{y} }{\partial y} = - {4\pi } \frac{\partial
\rho }{\partial t} +\varepsilon {A_{x}}{E_{y} }
\label{e24}
\end{equation}
\begin{equation}
- \frac{\partial B_z}{\partial x} = 4 \pi j_y + \epsilon
\frac{\partial E_y}{\partial t}\label{e23a}
\end{equation}
\begin{equation}
\frac{\partial B_z}{\partial y} = 4 \pi j_x + \epsilon
\frac{\partial E_x}{\partial t}\label{e24b}
\end{equation}
As some of these equatiions are time dependant, we are dealing with non-steady fields which give radiation.\\
This clearly brings out the extra electromagnetic effects. Because of (\ref{B}) there appears a magnetic field as was shown by the author and Saito \cite{bgsnc118b,saito}. We have in fact the equation
\begin{equation}
Bl^{2} =hc/e\label{e36}
\end{equation}
This clearly can be smoothly carried over to Graphene, keeping in mind the somewhat different values for the constants like $\nu_F$ and $l$. In fact we would have in this case
$$Bl^{2} =h \nu_F /e.$$

The energy in the above is given by
$$\mbox{Energy} \, = \, \pm \nu_F | \vec{p}|$$
The positive sign denotes conduction and the negative sign valence particles, the analogues of particles and antiparticles.\\
The analogy with high energy physics, particularly in the Cini-Toushek regime is very strong (Cf.ref.\cite{bgsijtp}). There too, we encounter a massless scenario. In fact at very high energies we have \cite{bgsijtp}
\begin{equation}
H \psi = \frac{\vec{\alpha} \cdot \vec{p}}{|p|} E(p)\label{C}
\end{equation}
which resembles the massless (\ref{A}) of the paper. Here in (\ref{C}) we have
\begin{equation}
\alpha^k = \left(\begin{array}{ll} 0 \quad \sigma^k\\
\sigma^k \quad 0\end{array}\right) \quad \quad \beta = \left(\begin{array}{ll} I \quad 0\\
0 \quad -I\end{array}\right)\label{24}
\end{equation}
\begin{equation}
\gamma^0 = \beta\label{26}
\end{equation}
This can be readily generalized to the neutrino equation. Importantly we have because of (\ref{B}) as discussed in the literature, the so called Snyder-Sidharth dispersion relation
\begin{equation}
E^2 = p^2 + m^2 + \alpha \frac{l^2}{\hbar^2} p^4\label{X}
\end{equation}
For Fermions $\alpha$ in (\ref{X}) is positive showing an extra contribution to the energy.\\
However there are differences with the usual Dirac Theory -- here we do not encounter Lorentz invariance and finally $\nu_F$ is not the velocity of light, rather its analogue.\\
We can see that Graphene will be a test bed in some interesting situations. The author had already argued several years ago \cite{fract1,fract2} that for nearly monoenergetic Fermions or even Bosons there would be a loss of dimensionality and the collection would behave as if it were in two dimensions. This immediately mimics the two dimensional feature.\\
Our starting point is the well known formula for the occupation number of a
Fermion gas\cite{huang}
\begin{equation}
\bar n_p = \frac{1}{z^{-1}e^{bE_p}+1}\label{8je9}
\end{equation}
where, $z' \equiv \frac{\lambda^3}{v} \equiv \mu z \approx z$
because, here, as can be easily shown $\mu \approx 1,$
$$v = \frac{V}{N}, \lambda = \sqrt{\frac{2\pi \hbar^2}{m/b}}$$
\begin{equation}
b \equiv \left(\frac{1}{KT}\right), \quad \mbox{and} \quad \sum \bar
n_p = N\label{8je10}
\end{equation}
Let us consider in particular a collection of Fermions which is
somehow made nearly mono-energetic, that is, given by the
distribution,
\begin{equation}
n'_p = \delta (p - p_0)\bar n_p\label{8je11}
\end{equation}
where $\bar n_p$ is given by (\ref{8je9}).\\
This is not possible in general - here we consider a special
situation of a collection of mono-energetic particles in equilibrium
which is the idealization
of a contrived experimental set up.\\
By the usual formulation we have,
\begin{equation}
N = \frac{V}{\hbar^3} \int d\vec p n'_p = \frac{V}{\hbar^3} \int
\delta (p - p_0) 4\pi p^2\bar n_p dp = \frac{4\pi V}{\hbar^3} p^2_0
\frac{1}{z^{-1}e^{\theta}+1}\label{8je12}
\end{equation}
where $\theta \equiv bE_{p_0}$.\\
It must be noted that in (\ref{8je12}) there is a loss of dimension
in momentum space, due to the $\delta$ function in (\ref{8je11}).\\
Similarly, recently the author had pointed out that the neutrinos behaved as if they were a two dimensional collection. Indeed \cite{ijtp2013} one could expect this from the holographic principle. Equally the author (and A.D. Popova) had argued that the universe itself is asymptotically two dimensional \cite{bgspopova}.\\
Furthermore it has also been argued that not only does the universe mimic a Black Hole, but also that the Black Hole is a two dimensional object \cite{bgsbh,tduniv}. Indeed the interior of a Black Hole is in any case inaccessible and the two dimensionality follows from the area of the Black Hole which plays a central role in Black Hole Thermodynamics. The author had shown, in his analysis that the area of the Black Hole is given by
\begin{equation}
A = N l^2_p\label{Ax}
\end{equation}
For these Quantum Gravity considerations we have to deal with the Quantum of area \cite{baez,tduniv}. In other words we have to consider the Black Hole to be made up of $N$ Quanta of area. Thus we can get an opportunity to test these Quantum Gravity features in two dimensional surfaces such as graphene.\\
In the earlier communication \cite{bgsjsp1999} it was shown that in the one dimensional case, corresponding to nanotubes we would have
\begin{equation}
kT = \frac{3}{5} kT_F\label{Bx}
\end{equation}
where $T_F$ is the Fermi temperature. It can be seen that for the two dimensional case too $kT$ is very small. This is because using the well known formulae for two dimensions we have
\begin{equation}
kT = \frac{e\hbar \pi}{m \nu_F}\label{5}
\end{equation}
\begin{equation}
(kT)^3 = \frac{6e\hbar \nu_F}{\pi}\label{4b}
\end{equation}
Whence we have
\begin{equation}
(kT)^2 = 6 \cdot \nu^2_F \pi^2 m\label{6}
\end{equation}
Remembering that $\nu_F \sim 10^8$, we have even for a particle whose mass is that of an electron, from (\ref{6}) $kT$ is very small. For a comparison we have for the Fermi temperature,
$$kT_F = \frac{\hbar}{2} (z6\pi)^{1/3} \cdot \nu_F$$
Another conclusion which could have been anticipated is the following. We have from the above
\begin{equation}
\nu^2_F = \left(\frac{\hbar \pi}{m}\right)^2 \cdot \frac{1}{A}\label{z}
\end{equation}
where $A \sim l^2$ is the quantum of area. So we get
\begin{equation}
\frac{m^2 \nu^2_F}{\hbar^2} \cdot l^2 \sim 0 (1)\label{za}
\end{equation}
This is perfectly consistent with $\nu_F$ tending to the velocity of light $c$ and $h/m \nu_F$ tending to the Compton wavelength. In other words an infinite graphene sheet would give us back the usual spacetime of Relativity and Quantum Mechanics. In practise we could expect this for a very large sheet of graphene. In either case it turns out that whatever be the temperature, it is as if the ensemble behaves like a very low temperature gas. This leads to many possibilities, particularly about magnetism.\\
As pointed out above we can investigate magnetism and electromagnetism in this new non-commutative paradigm which throws up novel features including the Haas Van Alphen type effect \cite{bgshj2013}. In this case, the magnetization per unit volume, as is known, shows an oscillatory type behaviour.
\section{Discussion}
Fluctuations of the Zero Point Field have been widely studied. Based on this the author in 1997 predicted a contra model of the universe \cite{ijmpa,tduniv} in which there would be a small cosmological constant, that is an accelerating universe. In 1998 observations of Perlmutter, Reiss and Schmidt confirmed this scenario. Today we call this dark energy. A manifestation of this is a noncommutative spacetime given in (\ref{B}). This lead to the so called Snyder-Sidharth dispersion relation given in (\ref{X}). We would like to point out that the extra magnetic effect in equations like (\ref{e23}) (and the following) can be attributed to this Zero Point effect of noncommutativity as given in (\ref{e36}). Closely related is the Casimir effect which has been observed even in graphene \cite{fial1,fial2}. This is a Zero Point Field fluctuation effect. The Casimir energy in graphene is given by
\begin{equation}
\frac{Energy}{area} = \frac{\pi^2}{240} \cdot \frac{\hbar c}{a^3}\label{x1}
\end{equation}
The energy itself is given by
\begin{equation}
Energy \, = \left(\frac{\pi^2}{240}\right) \cdot \frac{\hbar c}{a}\label{y1}
\end{equation}
where we consider the area to be $\sim a^2$.\\
If following Wheeler \cite{mwt} we consider directly ground state oscillators of the Zero Point Field, we can deduce that
$$Energy \, \sim \hbar c/a$$
resembling (\ref{y1}). Similarly if we take the extra term in the dispersion relation (\ref{X}), it is easy to show that this also has the same form. All this is hardly surprising because they are all manifestations of fluctuations in the Quantum Vacuum.\\
It must be mentioned that the Casimir effect in graphene has been observed. What is interesting is that a group of scientists from MIT, Harvard University, Oak Ridge National Laboratory and other Universities have used this Zero Point energy for a compact integrated silicon chip. Clearly the same would be possible for graphene too particularly in the context of Quantum Computers: The "Spin" up and down being the qubits \cite{zao}.\\
To proceed further we invoke (\ref{e36}) and the well known result for a coil
\begin{equation}
\imath = \frac{NBA}{R\Delta t}\label{HeE}
\end{equation}
where $N$ is the number of turns, $A$ is the area and $R$ the resistance. Use of (\ref{e36}) in (\ref{HeE}) now gives
\begin{equation}
\imath \approx \frac{NA}{R} \cdot \frac{e}{l^2\tau}\label{HeF}
\end{equation}
Whatever be $N$, if we think of a coil made up of nanotubes or graphene, remembering that $l$ is small and so is the resistance (\ref{HeF}) would be observable, like indeed (\ref{e36}).\\
Further observing that nanotubes and graphene can harbour fast moving Fermions (including neutrons) and of course carbon, we have all the ingredients for manipulating a version of table top fusion possibly using the bosonization of fermions property. In this case we use an equation like (\ref{8je12}) and preceding consideration \cite{tduniv,bgsjsp1999}.\\
To proceed, in this case, $kT = <E_p> \approx E_{p}$ so that,
$\theta \approx 1$.
But we can continue without giving $\theta$ any specific value.\\
Using the expressions for $v$ and $z$ given in (\ref{8je10}) in
(\ref{8je11}), we get
$$(z^{-1} e^\theta + 1) = (4\pi )^{5/2} \frac{z^{'-1}}{p_0};\mbox{whence}$$
\begin{equation}
z^{'-1}A\equiv z^{'-1}\left(\frac{(4\pi )^{5/2}}{p_0} -
e^\theta\right) = 1,\label{8je14}
\end{equation}
where we use the fact that in (\ref{8je10}), $\mu \approx 1$ as can
be
easily deduced.\\
A number of conclusions can be drawn from (\ref{8je14}). For
example, if,
$$A \approx 1, i.e.,$$
\begin{equation}
p_0 \approx \frac{(4\pi )^{5/2}}{1+e}\label{8je15}
\end{equation}
where $A$ is given in (\ref{8je14}), then $z' \approx 1$.
Remembering that in (\ref{8je10}), $\lambda$ is of the order of the
de Broglie wave length and $v$ is the average volume occupied per
particle, this means that the gas gets very densely packed for
momenta given by (\ref{8je15}). Infact for a Bose gas, as is well
known, this is the condition for Bose-Einstein condensation at the
level
$p = 0$ (cf.ref.\cite{huang}).\\
In any case there is an anomalous behaviour of the Fermions.

\end{document}